\documentclass[%
 reprint,
superscriptaddress,
showpacs, showkeys,
footinbib,
 amsmath,amssymb,
 aps,
prstab,
]{revtex4-1}
\usepackage{multirow}
\usepackage{ulem}
\usepackage{longtable}
\usepackage{array}
\usepackage{graphicx}
\usepackage{dcolumn}
\usepackage{bm}
\usepackage{hyperref}
\usepackage{enumerate}
\usepackage{color}

\begin{document}

\preprint{APS/123-QED}

\title{Non-volatile, reversible	metal-insulator	  transition in	
  oxide	interfaces controlled	by gate voltage	and light}%

\author{Mian Akif Safeen}
\thanks {These authors equally contributed to the work}
\affiliation{Department of Physics, Abdul Wali Khan University, Mardan 23200, Pakistan}
\affiliation{CNR-SPIN UOS Napoli} 
\affiliation{Dipartimento di Fisica  ``E. Pancini", Universit{\'a} di Napoli ``Federico II", Compl. Univ. di Monte S. Angelo, Via Cintia, I-80126 Napoli, Italy}%
\author{Musa Mutlu Can}
\thanks {These authors equally contributed to the work}
\affiliation{Department of Physics, Faculty of Science, Istanbul University, Vezneciler, 34314, Istanbul, T{\"u}rkiye}\affiliation{CNR-SPIN UOS Napoli} 
\author{Amit Khare}
 \affiliation{Department of Physics, Indian Institute of Science Education and Research, Bhopal - 462 066 (India).}\affiliation{CNR-SPIN UOS Napoli}
\author{Emiliano Di Gennaro}
\affiliation{Dipartimento di Fisica  ``E. Pancini", Universit{\'a} di Napoli ``Federico II", Compl. Univ. di Monte S. Angelo, Via Cintia, I-80126 Napoli, Italy}\affiliation{CNR-SPIN UOS Napoli} 
\author{Alessia Sambri}
\affiliation{NEST, Istituto Nanoscienze-CNR, Piazza San Silvestro 12, 56127 Pisa } \affiliation{CNR-SPIN UOS Napoli} 
\author{Antonio Leo}
\affiliation{Dipartimento di Fisica ``E. R. Caianiello" Universit{\'a} di Salerno, Fisciano, Italy} 
\author{N. Scopigno }
\affiliation{Dipartimento di Fisica, Universit\`a di Roma ``La Sapienza'', P.$^{le}$ Aldo Moro 5, 00185 Roma, Italy}%
\author{Umberto Scotti di Uccio}
 \affiliation{Dipartimento di Fisica  ``E. Pancini", Universit{\'a} di Napoli ``Federico II", Compl. Univ. di Monte S. Angelo, Via Cintia, I-80126 Napoli, Italy}\affiliation{CNR-SPIN UOS Napoli}
\author{Fabio Miletto Granozio }
\email{Corresponding author: fabio.miletto@spin.cnr.it}
\affiliation{CNR-SPIN UOS Napoli} \affiliation{Dipartimento di Fisica  ``E. Pancini", Universit{\'a} di Napoli ``Federico II", Compl. Univ. di Monte S. Angelo, Via Cintia, I-80126 Napoli, Italy}%
\date{\today}
\begin{abstract}
The field-effect-induced modulation of transport properties of 2-dimensional electron gases residing at the LaAlO$_3$/SrTiO$_3$ and LaGaO$_3$/SrTiO$_3$ interfaces has been investigated in a back-gate configuration. Both samples with crystalline and with amorphous overlayers have been considered. We show that the ``na{\"i}ve" standard scenario, in which the back electrode and the 2-dimensional electron gas are simply modeled as capacitor plates, dramatically fails in describing the observed phenomenology. Anomalies appearing after the first low-temperature application of a positive gate bias, and causing a non-volatile perturbation of sample properties, are observed in all our samples. Such anomalies are shown to drive low-carrier density samples to a persistent insulating state. Recovery of the pristine metallic state can be either obtained by a long room-temperature field annealing, or, instantaneously, by a relatively modest dose of visible-range photons. Illumination causes a sudden collapse of the electron system back to the metallic ground state, with a resistivity drop exceeding four orders of magnitude.  The data are discussed and interpreted on the base of the analogy with floating-gate MOSFET devices, which sheds a new light on the effects of back-gating on oxide-based 2-dimensional electron gases. A more formal approach, allowing for a semi-quantitative estimate of the relevant surface carrier densities for different samples and under different back-gate voltages,  is proposed in the Appendix .

\end{abstract}

\maketitle


\section{\label{sec:Intro}Introduction}
The electron density in 2-dimensional electron gases (2DEGs) at oxide interfaces can be very effectively modulated by electric field effect \cite{thiel_tunable_2006}. Such possibility has been successfully exploited to tune the properties of the LaAlO$_3$/SrTiO$_3$ (LAO/STO) system and of its variants, e.g. for spanning the superconducting \cite{caviglia_electric_2008} magnetic \cite{stornaiuolo_tunable_2016}, and spin-orbital \cite{ben_shalom_tuning_2010,caviglia_tunable_2010} phase diagrams. The effect of an applied electric field on the transport properties has been  typically interpreted in terms of a standard field effect transistor model, in which charge is transferred between the 2D electron gas (2DEG) and the gate electrode as in a capacitor charge/discharge process. Further measurements highlighted a quite complex response of the transport properties of the 2DEG as a function of gate voltage. Bell et al., \cite{bell_dominant_2009} and more recently Biscaras et al. \cite{biscaras_limit_2014} demonstrated that anomalous effects, well beyond the capacitive charge-discharge of a field-effect device,  occur during and after the  application of a positive back-gate (BG) voltage. Such BG bias perturbs in fact the properties of the 2DEG, bringing it into a metastable state with a conductivity reduction up to about a factor two. The effect is qualitatively interpreted in \cite{biscaras_limit_2014} in terms of charges flowing away from the quantum well during a positive gate polarization. Beyond this, no attempt of microscopic and electrostatic analysis has been proposed so far and no method to switch the system back from the metastable to the pristine state has been found.

In this paper, the persistent effect of the first positive polarization is on three suitably chosen characteristic samples with different initial carrier density is analyzed and discussed. We show that a ``colossal" switch with a resistive ratio $R_{OFF}/R_{ON}>10^4$ can be obtained in a specific class of oxide interfaces (represented by sample \textbf{C} in Table \ref{tab1}). The nature of the metastable highly resistive state is addressed in detail, exploiting a strong analogy with the retention state of floating-gate (FG) MOSFET device. We argue that the formation of such insulating retention state is related to a negative space charge region formed by electrons that are localized in shallow traps in proximity of the interface, that electrostatically gate the quantum well. We show that, in full analogy with the case of the semiconducting FG-based EPROM devices, exposure to light can be employed to detrap the electrons destabilizing the metastable retention state. Illumination causes the instantaneous collapse of the electron system back to the metallic ground state, with a resistivity drop exceeding in Sample \textbf{C} four orders of magnitude.
\begin{table*}[t]
\centering
\begin{tabular}{cccccccc}
\hline
\multirow{2}{*}{Name} & \multirow{2}{*}{System} & STO         & Thickness & PO$_2$    & R$_{Sheet}@290K$           & R$_{290K}$/ & \textbf{n$_{Sheet}@290K$}   \\
                      &                         & Orientation & (nm)      & (mbar)    & (k$\Omega$/$\square$) & R$_{10K}$   & \textbf{(cm$^{-2}$)}  \\ \hline
A                     & c-LAO/STO               & (001)       & 4         & 10$^{-3}$ & 10                    & 18          & \textbf{1.1x10$^{14}$} \\
B                     & c-LAO/STO               & (001)       & 4         & 10$^{-2}$ & 14                    & 25          & \textbf{8.2x10$^{13}$} \\
C                     & a-LGO/STO               & (001)       & 4         & 10$^{-2}$ & 68                    & 168         & \textbf{2.1x10$^{13}$} \\ \hline
\end{tabular}
\caption{Samples fabrication details and transport properties.}
\label{tab1}
\end{table*}

\section{\label{sec:met}Methods and samples.}
Our experiments have been performed on several heterostructures, including ``standard" (001) oriented, crystalline LAO/STO interfaces, as first reported by Ohtomo and Hwang \cite{ohtomo_high-mobility_2004},  (110) oriented crystalline LAO/STO interfaces as first reported by Herranz et al. \cite{herranz_high_2012}, and LaGaO$_3$/SrTiO$_3$ (LGO/STO) interfaces, first studied by some of the authors\cite{perna_conducting_2010}. For LGO/STO interfaces, also the ``amorphous" variant, following the results reported in \citep{chen_metallic_2011} using different amorphous overlayers, was investigated. 
All films were grown by RHEED-assisted pulsed laser deposition on TiO$_2$-terminated STO substrates. Relatively high-pressure growth conditions, ranging between 10$^{-3}$ and 10$^{-2}$ mbar O$_2$, were employed for the growth. Crystalline samples were grown at 730$^\circ$C and hence slowly cooled in the process gas. Amorphous samples were grown at room temperature. The electrical transport measurements were performed in a close-cycle refrigerating system. The gate voltage was applied on the back side of a 0.5 mm thick STO substrate. Samples were kept in dark for at least 12 hours before the first measurement. All the transport characterization was carefully performed in dark conditions, except for the measurements that were specifically designed to be made under light. Four-contact Van der Pauw measurements were performed with a typical polarization current of 10 $\mu$A.   

\section{Results and Discussion}
Data from three representative samples of our overall batch are reported and discussed in the following. Samples \textbf{A} and  \textbf{B}  are crystalline LAO/STO heterostructures grown on (001) TiO$_2$-terminated STO.  Both are referred to as c-LAO/STO interfaces, where $c$ stands for crystalline. Sample \textbf{C} is a LGO/STO interface  having an amorphous LGO overlayer. It is referred to as a-LGO/STO interface, where $a$ stands for amorphous. All samples were metallic. This is the first report, to our knowledge, of a metallic  a-LGO/STO interface.
\begin{figure}
\includegraphics[width=0.48\textwidth]{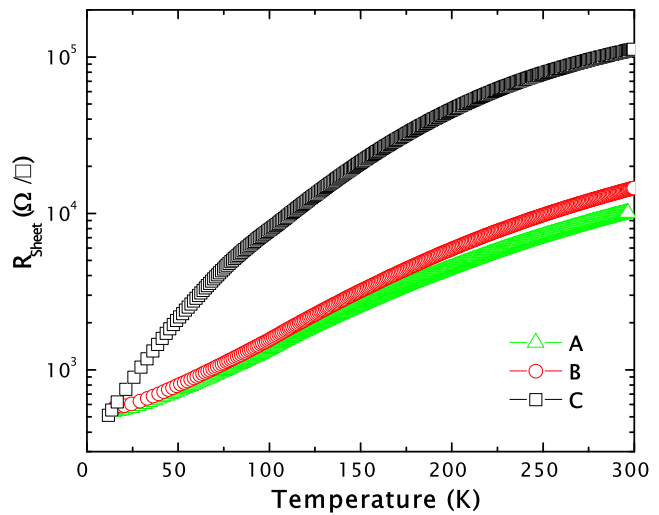} 
\caption{R$_{Sheet}$(T) curves for the three samples described in Table \ref{tab1}}
\label{fig1}
\end{figure}

 All R(T) curves are reported in Fig.\ref{fig1}. The main transport properties of the three samples are summarized in Table \ref{tab1}. Interestingly, the a-LGO/STO sample has a lower carrier density, a higher room temperature resistivity and a higher resistive ratio with respect to the c-LAO/STO samples reported in this work. Such behavior, as checked in similar samples, is typical of this specific heterostructure in the given growth conditions.
 
Data presentation will proceed as follows. We will start by analyzing sample  \textbf{A}, showing a back-gate voltage response similar to the one reported in \cite{biscaras_limit_2014} and will show that such previously observed behavior can be understood by resorting to an analogy with a class of current semiconductor devices. The sample response is presented within each step of the back-gate voltage waveform, $V_G(t)$,  and discussed in the light of the analogy above. The same concepts are then adopted to present the more anomalous responses of sample \textbf{B}, whose intermediate behavior ``bridges" sample  \textbf{A} to \textbf{C}. Finally, the colossal switch of \textbf{C} is addressed and the effect of light as a resetting tool, able to abruptly drive the system back to the pristine state, is presented. Further details about the properties of our samples are reported in \cite{mynote5}.
\subsubsection{Sample A}
\begin{figure}
\includegraphics[width=0.48\textwidth]{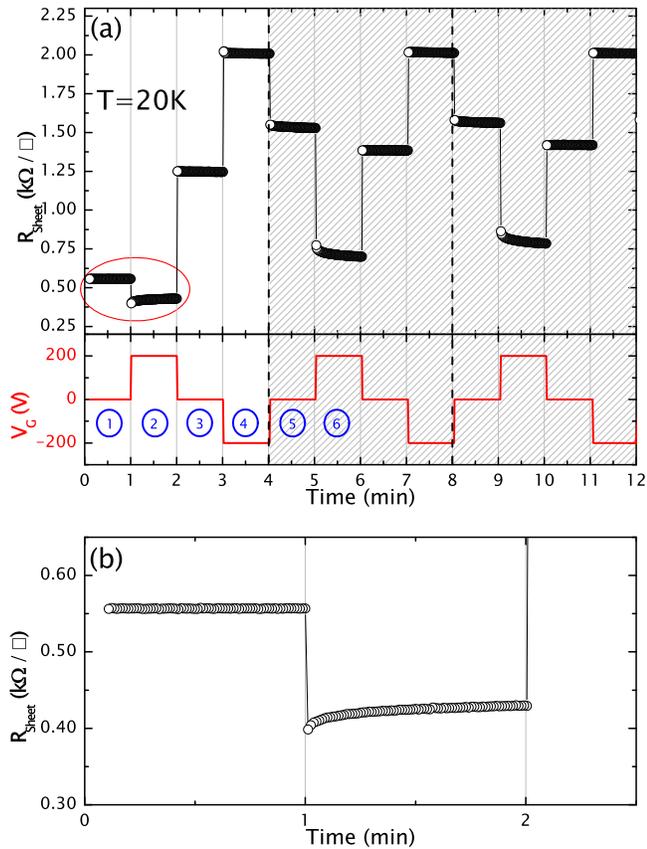} 
\caption{(a) Temporal evolution of R$_{Sheet}$ for sample \textbf{A} at 20K (upper panel) modulated by the applied gate voltage (V$_G$) (lower panel). Dashed vertical lines separate the three different measured periods of our square waveform. The unshaded area indicates the first period. The first six quarter-cycles, or phases, of the waveform are labeled with blue circles. (b) A detailed view of the FPP effect obtained by magnifying the region circled in red in the upper panel of (a), i.e. phase n. 1 and 2 (a).}
\label{fig2}
\end{figure}

The plot shown in Fig. \ref{fig2} describes the effect of the first gate voltage cycle on the pristine sample \textbf{A} at 20K. We intend here by ``pristine" a sample that was never employed in a field-effect experiment before. The slow square back-gate voltage waveform $V_G(t)$ shown in Fig. \ref{fig2}a, lower panel, with a period of 4 min, was applied. The measured response differs by several features from the one expected within the typical capacitive model of a field-effect device. Such features include:
\begin{enumerate}[i)]
\item an anomalous first cycle;
\item the presence of slow dynamics with time constants well beyond typical RC constants;
\item a hysteretic behavior, evident in the fact that the zero-bias sheet resistance value is different after a positive and a negative bias;
\end{enumerate}
The discussion of the last two features, visible in the shaded portion of Fig. \ref{fig2}a, is postponed to future works. We will focus here our attention on the effect on the 2DEG properties of the  ``first positive polarization" (FPP), i.e. on the phenomena taking place during the transition between the first and the second quarter-cycle of the square waveform, as indicated in the lower panel of Fig. \ref{fig2}a. The plot of the sheet resistance recorded vs. time during the FPP is magnified in Fig. \ref{fig2}b.
It is shown that the resistance drop obtained during the FPP is relatively small, compared to the R$_{Sheet}$ variations obtained during the subsequent positive or negative voltage steps.  A partial and slow recovery of the resistance takes place while the +200 V gate voltage is applied. More importantly, when the gate voltage is reduced to zero after the FPP, the sample resistivity increases by over a factor two with respect to the pristine sample. During the following cycles, the initial resistance value is never recovered. When measurements similar to the one reported in Fig. 2 are repeated at different temperatures and different voltage values, the resistivity grossly scales with the STO dielectric constant $\epsilon$ and with the applied electric field, showing a threshold behavior as a function of the $\epsilon E$ product. Negligible effects are obtained above 50K for the given values of the applied electric fields. A behavior similar to the one reported in Fig.2 was observed in the paper by Biscaras et al., which showed that the sheet resistance change was directly related to a carrier density change and attributed such effect to a flow of electrons creeping over the edge and simply ``disappearing" from the quantum well. 

On the base of the measurements reported in our work, we will show that the electrons pouring out of the quantum well should not simply regarded as ``lost carriers". While lost to conduction, the localized electrons effectively reshape the interface potential profile with their space charge. As a consequence, the behavior of our samples is dominated by the electrostatic interplay between two distinct ``electron reservoirs", mobile carriers and localized charges, that interact electrostatically, while dynamically exchanging electrons under the external electric field and light. 

Generally speaking, the most straightforward possible explanation for the presence of a hysteresis in a R-V$_{G}$ cycle is a ferroelectric behavior of the gate barrier. The analysis of the first cycle in Fig. \ref{fig2}a (divided in the four quarter-cycles named 1, 2, 3, 4) shows nevertheless that this explanation can not be applied to this specific case. After the first positive polarization (2$^{nd}$ quarter-cycle), which increases the electron density in the 2DEG, the presence of a remnant polarization, alone, would maintain the sample in a high-carrier-density, low-resistivity state. This is in contrast with the fact that the R$_{Sheet}$ during the 3$^{rd}$ quarter-cycle is higher than in the pristine state (1$^{st}$ quarter-cycle). In order to understand such peculiar behavior, a more accurate look at the band diagram of our system is needed. An accurate experimental analysis of the irreversible behavior taking place under positive polarization has been reported by  \citep{biscaras_limit_2014}. We will show that an intuitive understanding of the whole phenomenology can be achieved by comparing the band diagram of our system to the one of conventional semiconductor devices. A more formal approach is reported in Appendix A.

\begin{figure}
\includegraphics[width=0.48\textwidth]{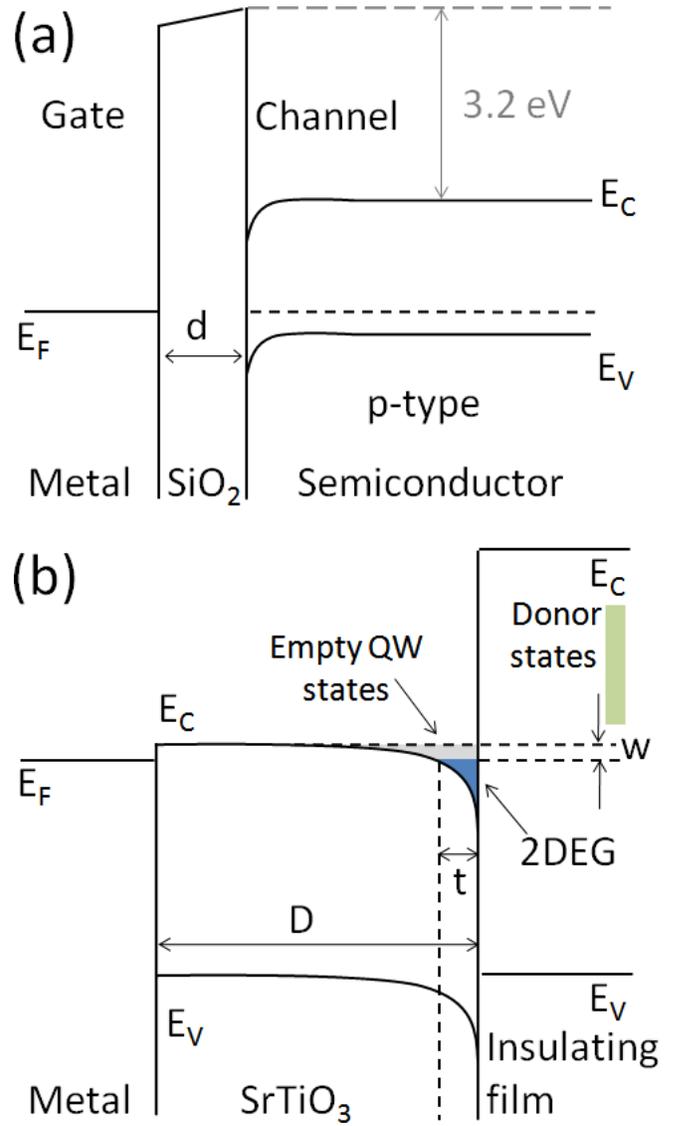} 
\caption{Energy band diagrams for a SiO$_2$-MOS capacitor (a) and a back-gated LAO/STO interface (b). The height $w$ of the small and smooth potential step separating the 2DEG Fermi Level from the
lowest states of the bulk STO conduction band is shown, together with the 2DEG spatial width $t$. The presence of a possible band-bending within the insulating film, as in the polar catastrophe scenario, is irrelevant for our arguments and is hence ignored. The presence of ionized donor states in the film above the Fermi level is represented by a green vertical line. Note that all horizontal distances are not in scale, being $D\gg d$ and $D\gg t$.}
\label{fig3}
\end{figure}

Let's start by considering as a first possible analogous of our system one of the main building blocks of current semiconductor technology, i.e., a MOS capacitor with a SiO$_2$ barrier, shown in Fig. \ref{fig3}a. The barrier height, i.e. the energy difference between the conduction band minimum within SiO$_2$ and within the doped-Si electrodes, is of the order of 3.2 eV. Although such value is influenced by the gate voltage, it's relative variation is small. The oxide region remains therefore practically inaccessible to the carries present in the channel and the current losses through the barrier are mainly related to tunneling.  

We now observe that in a back-gated oxide 2DEG, the band diagram at the channel/barrier interface (i.e. the 2DEG/STO interface) is substantially different. The sketch reported in Fig. \ref{fig3}b   has been taken from \cite{treske_universal_2015}. A small and smooth potential step separates the 2DEG Fermi level from the lowest empty states of bulk STO. Estimates of the depth of the quantum well of the order of 0.3 - 0.6 V have been reported \citep{huang_mapping_2012,popovic_origin_2008}. 

The main effect of the application of a back-gate voltage is to modify the energy landscape in the proximity of the 2DEG, through a combination of the Fermi level variation related to the increased filling and the added slope related to the external field \cite{scopigno_phase_2016}.

For a positive gate voltage, the surface charge induced by field effect can easily fill the few empty confined states present in the quantum well, inducing a flux of free extra electrons to creep out and migrate, following the weak potential gradient. Our experiment tells that such charges cause a non-volatile memory effect, pushing the device towards a high-resistance state. This suggests that, as will be shown in the following,  the correct analogous of our system in semiconductor technology has to be searched, rather than in a MOS capacitor, in a more complex, non-volatile, memory device. 

In order to describe the observed behavior under the square potential backgate form $V_G(t)$, we will address the sample response  both at a qualitative level, based on the analogy with commercial semiconductor devices, and at a semi-quantitative level, resorting to a surprisingly simple electrostatic approach.  Let’s briefly describe the basic concepts of our model. 

Following \cite{scopigno_phase_2016}, we impose a charge neutrality condition for our samples by assuming that a positive charge distribution  $\sigma_{\!_{DON}}$ of ionized donors is located outside STO, across the interface on the right, on the layer side. This assumption is in agreement with a standard ER electronic reconstruction scenario \cite{ohtomo_high-mobility_2004,nakagawa_why_2006}, with a defect-assisted ER scenario\cite{bristowe_surface_2011,yu_polarity-induced_2014} and with other scenarios in which, even in absence of polarity, the donor states in the amorphous overlayer dope the quantum well \cite{chen_extreme_2015}. The   $\sigma_{\!_{DON}}$  distribution is depicted in Fig. 3b. An equal number of electronic charges dopes the interface at the STO side, in the absence of a back-gate voltage. The electrostatic attraction creates a QW, hosting a number of bound states equal to $\sigma_{\!_{DON}}$/e \cite{scopigno_phase_2016}. In agreement with established experimental evidences, showing that the electron densities depend drastically on deposition conditions, the $\sigma_{\!_{DON}}$ density is not fixed to 0.5 electrons per in-plane unit cell, but is a specific sample-dependent property.
In the proximity of the interface, localized acceptor states (LAS), able to trap electrons are present \cite{scopigno_phase_2016,ristic_photodoping_2012}. 
Let’s now define the following surface charge densities (SCD): 

\begin{longtable}{p{0.15\linewidth} p{0.8\linewidth}}
$\sigma_{\!_{DON}}$   & the fixed positive SCD due to charges localized in the layer (in our case $LAO$ or $LGO$) on the right side of the interface, shown in the right side of Fig. \ref{fig3}b, guaranteeing the the overall charge neutrality of the system \cite{scopigno_phase_2016}.\\
$\sigma_{\!_{QW}}$    & the total SCD that can be hosted in the QW, sum of the states depicted in blue and in gray in Fig. \ref{fig3}b:  $\sigma_{\!_{QW}}$=$\sigma_{\!_{2DEG}}$+$\sigma_{\!_{EMPTY}}$. $\sigma_{\!_{QW}}$   depends on $V_G$, similarly to the other SCDs defined below, because of the additional constant potential slope introduced by  $V_G$ , that perturbs the QW profile.  \\
$\sigma_{\!_{2DEG}}$  & the negative SCD of mobile electrons in the quantum well, depicted in blue in Fig. \ref{fig3}b.\\
$\sigma_{\!_{EMPTY}}$ & the negative SCD corresponding  to states available for mobile electrons confined in the quantum well, depicted in gray in Fig. \ref{fig3}b, that are empty in the pristine state, i.e. before application of a gate voltage.\\
$\sigma_{\!_{LAS}}$   & the negative SCD of trapped electrons in the localized acceptor states. Here we are focusing our attention on states that are external to the quantum well and are totally or partially empty in equilibrium conditions. Such states are  shown Figs. \ref{fig4}b,d and f. States of the same chemical origin, i.e. produced by the same kind of defects, but located closer to the interface, will have a lower energy due to the bending of the energy profile and will therefore be full in equilibrium conditions. Such filled states play no role in the following arguments and are, for the sake of graphical clarity, omitted in Fig. \ref{fig4}b,d and f.  \\
$\sigma_{\!_{IND}}$   & the negative SCD induced in the interface by a positive 200V bias. An opposite, positive SCD (-$\sigma_{\!_{IND}}$) will be induced at the same time at the gate. The sign of $\sigma_{\!_{IND}}$ will be reversed for a negative bias. $\sigma_{\!_{IND}}$ is the only quantity that is not sample-dependent, assuming that all STO substrates are identical. \\
\end{longtable}
On the base of the self-consistent calculation of the electrostatic potential reported in  \cite{scopigno_phase_2016}, it was shown that the energy defined as $w$ in Fig. \ref{fig3}b should be zero in absence of a population of localized charges $\sigma_{\!_{LAS}}$ near the interface. This observation is a quite straightforward  consequence of the one-dimensional Poisson equation. This is readily recognized by observing that the CB profile in STO, outside the 2DEG (i.e. in absence of a space charge), can only have a constant slope, fixed by the external gating $V_G$. 
  
We will show in the following that $\sigma_{\!_{EMPTY}}$ is pretty small in the pristine samples with respect to the amount of charge $\sigma_{\!_{IND}}$  that is transferred by field effect. We will also show that, due to the above described difference between the band diagrams reported in Fig. \ref{fig3}a and Fig. \ref{fig3}b, and to the presence of localized acceptor states near the interface, the correct analogous of our system is given not by the MOS capacitor shown in Fig. \ref{fig3}a,  but by the floating-gate MOSFET device described in Fig. \ref{fig4}a, c, e. 

In Appendix \ref{sec:Model} we describe the behavior of each sample during the quarter-cycles, within the general framework set above, resorting to a number of very reasonable and straightforward assumptions. Furthermore, a semi-quantitative evaluation of all the above described SCDs will be provided for each sample in each step of the $V_G(t)$ waveform.

As shown in Fig. \ref{fig4}a,c the functioning mechanism of FG MOSFETs is based on the possibility to transfer carriers from the channel to the FG states (Fig. \ref{fig4}c) through the thin separating oxide barrier $d_2$, under positive bias. The electric field caused by the negatively charged FG keeps the MOSFET in an OFF state, the so-called retention state, even when the control gate is unbiased (Fig. \ref{fig4}e). It is easily anticipated that carriers bubbling out of the 2DEG  can have the same effect, if they can accumulate in localized acceptor states (LAS) present in the STO barrier, such to create a space-charge region. We postulate the presence of such LAS now (their origin will be discussed later) and depict them graphically in Figs. \ref{fig4}b,d,f as a series of shallow potential wells. Their energy depth can be estimated of the order of $10^{-1}$ eV.
Let’s now exploit the analogy with FG devices to address our data. 
\begin{figure*}
\includegraphics[width=0.90\textwidth]{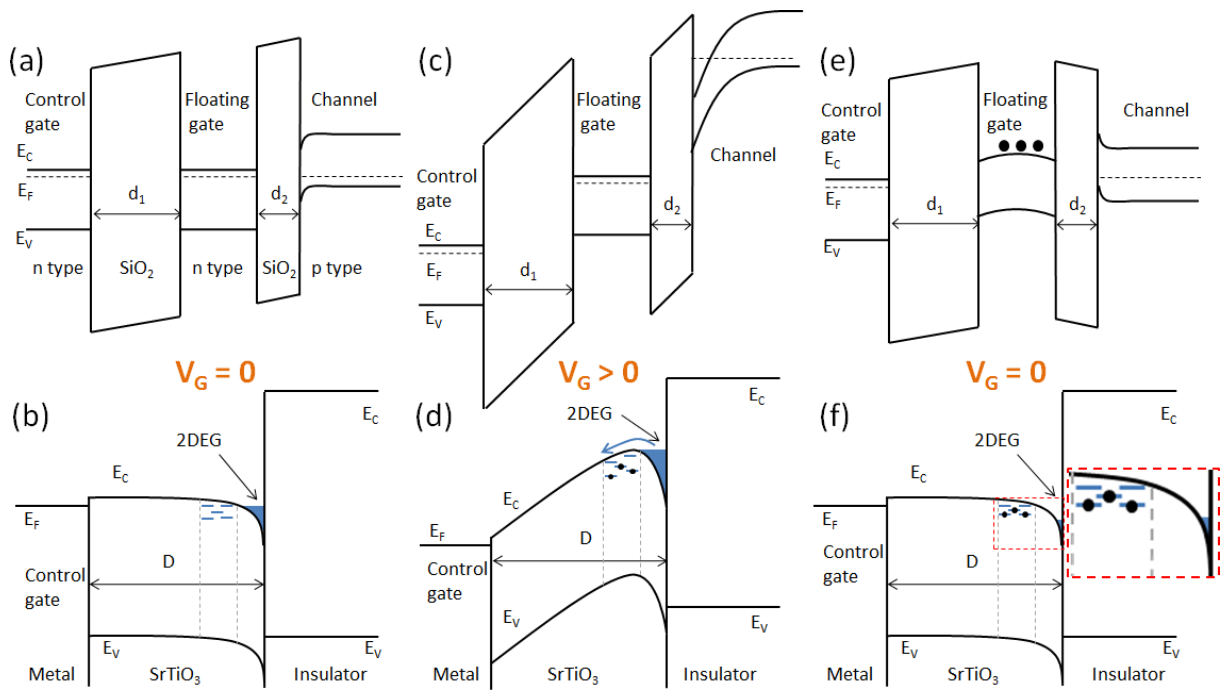} 
\caption{Comparison between the band diagram of a FG MOSFET device and a back-gated oxide 2DEG. (a) and (b) Pristine state of both the semiconductor and oxide devices with short-circuited control gate and channel; (c) and (d) positive control gate polarization inducing the retention state; (e) and (f): retention state at zero external bias. The presence of localized acceptor states (LAS) in STO, in vicinity of the 2DEG, is postulated and will be discussed in the following. The blue arrow in (d) represents the creeping out mechanism. Note that all horizontal distances are not in scale and in particular D$\gg$d$_1\gg$d$_2$}
\label{fig4}
\end{figure*}

\textit{Phase 1}\\
During phase 1, no bias is applied. The sheet resistance value is 550$\Omega$/$\square$.\\

\textit{Phase 2}\\
As soon as $V_G$ is increased to 200V, this FPP pushes an extra sheet density of capacitively induced charges $\sigma_{\!_{IND}}$, adding to the native sheet density of the 2DEG $\sigma_{\!_{2DEG}}$ , towards the QW through the control electronics. $\sigma_{\!_{IND}}$ can be estimated of the order of 2-3 x 10$^{13}$ cm$^{-2}$ at 200V, 20K\cite{hirose_electric_2015}. The flow of $\sigma_{\!_{IND}}$ towards the QW during the positive ramp causes the resistance of the sample first to decrease and then to saturate quickly, as soon as, at a critical gate bias, the sheet density of states in the QW ($\sigma_{\!_{QW}}$) is filled. All extra charges induced above such bias, not finding available states in the QW, are pulled away from the interface by the electric field present in the barrier.  Fig. \ref{fig4}d shows that such excess electrons, trapped in the LAS, act similarly to the trapped electrons in a FG device, persistently changing the resistance of the sample. 
In the course of this second step, at constant positive $V_G$, a small damped upturn of the resistance of about 5\% takes place. As previously observed in \cite{biscaras_limit_2014}, the QW, previously brought to complete filling by the FPP (Fig. \ref{fig4}d), slowly looses the electrons lying, energywise, in proximity of the QW edge, for a  ``thickness" of the order of k$_B$T. Such electrons drift along the potential gradient and are eventually also trapped in the LAS. When a similar analysis as the one performed in \cite{biscaras_limit_2014} was applied, similar values for the characteristic escape time $t_E \approx 1s$ were found.\\

\textit{Phase 3}\\
When the BG bias is set back to zero, our interface is left in a high resistive retention state, in analogy to a FG MOSFET device (Fig. \ref{fig4}e). Such state is metastable, since at $V_G$=0 the empty quantum well states lye again below the trap states filled during the FPP. The sample is unable to recover the pristine ground state, unless special procedures are applied, as discussed later in this work.  The sample resistance is increased by about a factor two. The trapped electrons perturb the electrostatics of the system.  Trapped electrons effectively gate the 2DEG by inducing in it an opposite positive surface charge density, as can be inferred by charge conservation. The semi-quantitative model reported  in Appendix A provides, for sample A in the third step,  the estimate,  $\sigma_{\!_{2DEG}}=-\sigma_{\!_{DON}}-\sigma_{\!_{IND}} $ .  \\

\textit{Phase 4}\\
The sample resistance is further increased by about an extra factor two. In this configuration, the electrostatic effect of the negatively charged FG and BG  add to each other, doubling the effect of the a single negative gate. Hence we have:  $\sigma_{\!_{2DEG}}=-\sigma_{\!_{DON}}-2\sigma_{\!_{IND}} $ .  \\
Again, this result will be more formally derived in Appendix A.
\\

During the successive cycles, no further irreversible effect takes place. Once the excess electrons have been ``trimmed" during the FPP, in fact, the Fermi level never exceeds again the edge of the QW, even under positive bias. During the following steps, the electrons are  reversibly transferred between the back gate and the 2DEG through the control electronics without further transfer to/from the LAS. Let’s now analyze one by one the four quarter-cycles of the BG square wave on Samples B and C, and in particular the FPP effect.
\begin{figure}
\includegraphics[width=0.48\textwidth]{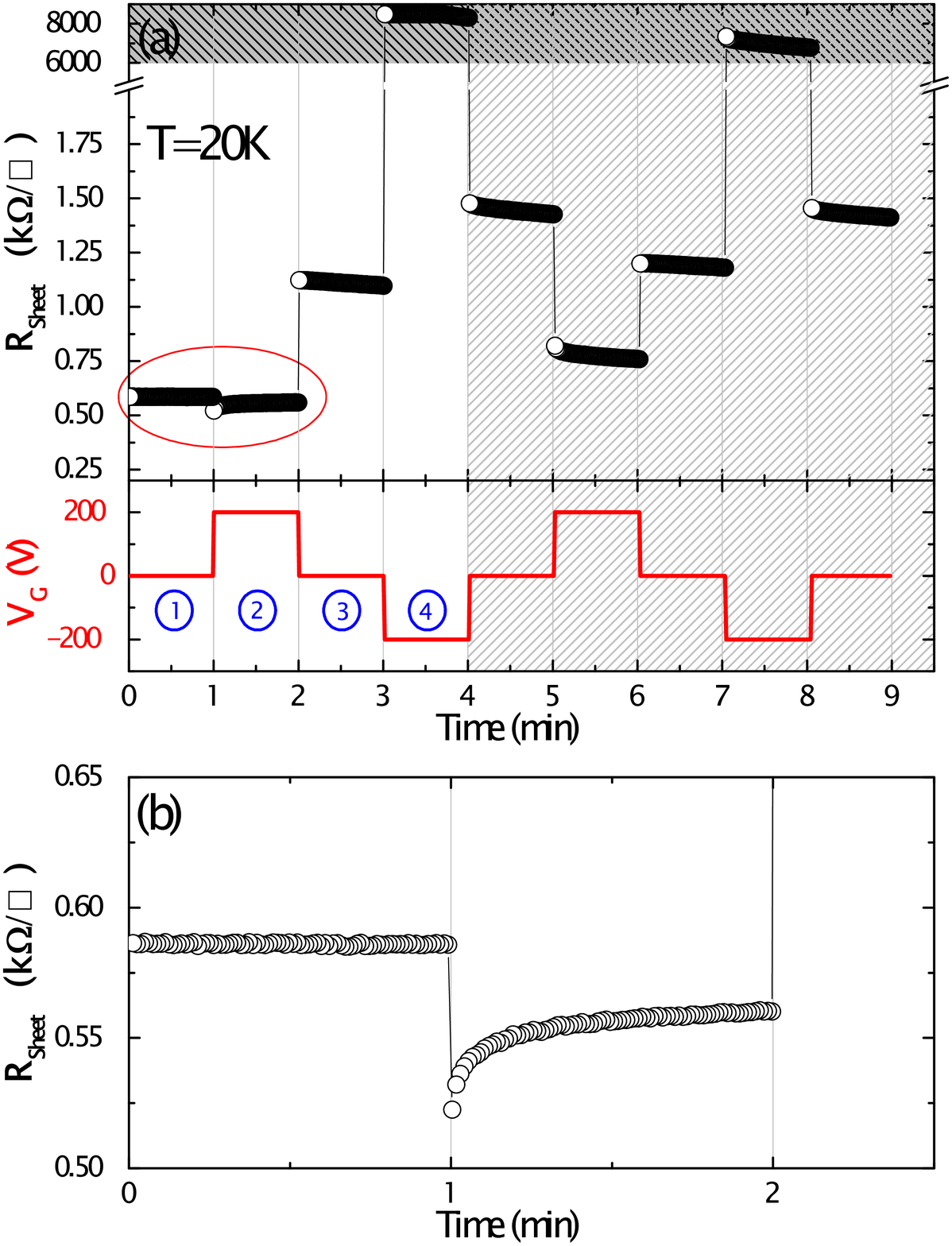} 
\caption{(a) Temporal evolution of R$_{Sheet}$ for sample \textbf{B} at 20K (upper panel) modulated by the applied gate voltage (V$_G$)(lower panel). The dark-gray area indicates the instrument compliance limit. The dark-gray area indicates the instrument compliance limit. The brighter area indicates the temporal region under analysis in this work, for sake of clarity split  in four phases according with the V$_G$ modulation and labeled with the blue circles. (b) A detailed view of the temporal behavior in the region indicated by the red circle in the upper panel  of (a).}
\label{fig5}
\end{figure}
\subsubsection{Sample B}
During Phase 1, at 20K, the sample has an R$_{Sheet}=580\Omega/\square$. The decrease of sheet resistance taking place during Phase 2 (Fig. \ref{fig5}) is pretty small. This suggests that the QW is, as in the previous case, very close to complete filling in the virgin state. Once again, the  second zero-bias state , in retention mode (Phase 3), has a resistance that is about a factor 2 higher than the virgin state. A quantitative difference with the case of Sample \textbf{A} is found in the negative polarization state (Phase 4). The negative BG bias here completely depletes the QW well of mobile charges, bringing the sample to an insulating state.
\begin{figure}
\includegraphics[width=0.48\textwidth]{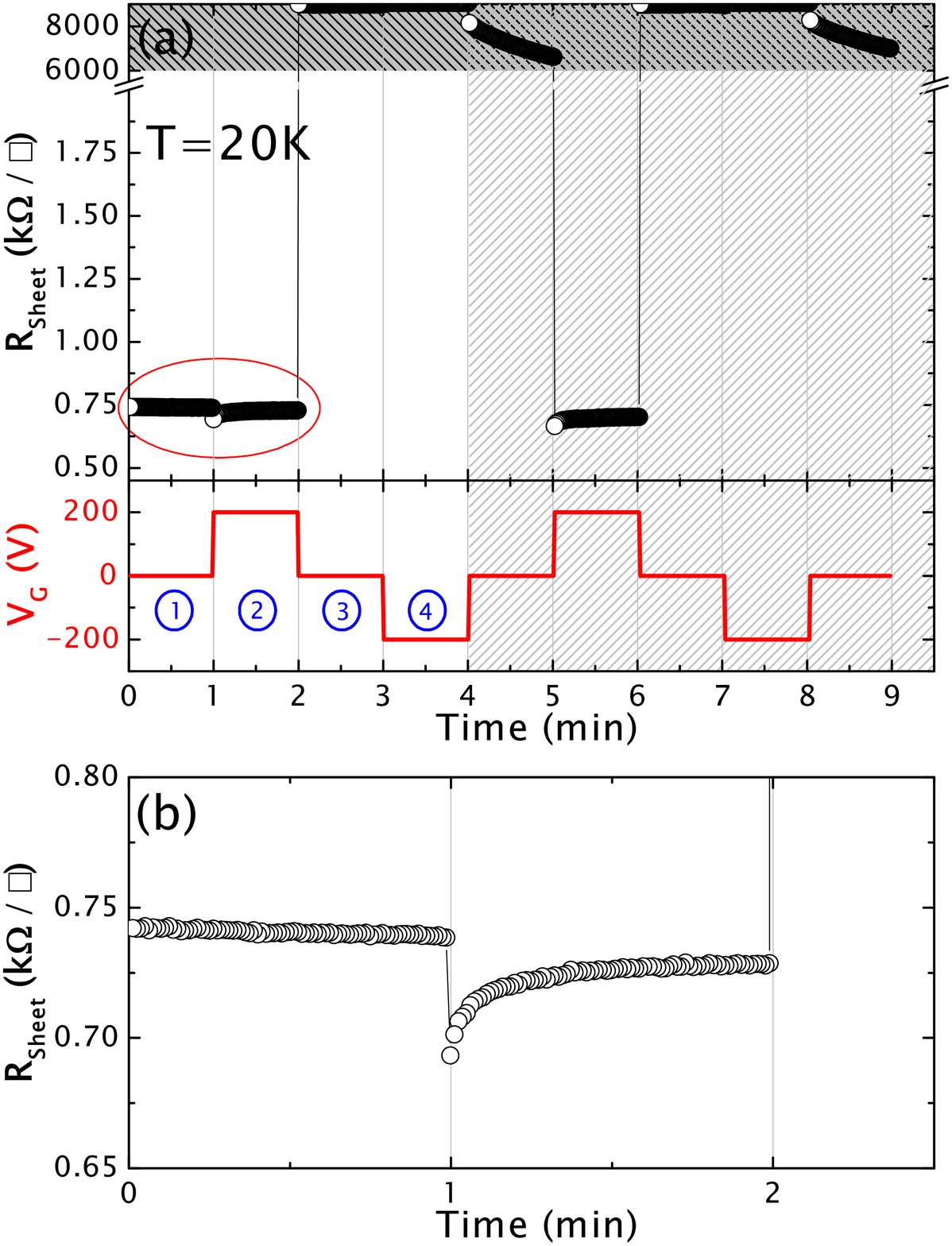} 
\caption{(a) Temporal evolution of R$_{Sheet}$ for sample \textbf{C} at 20K (upper panel) modulated by the applied gate voltage (V$_G$)(lower panel). The dark-gray area indicates the instrument compliance limit. The dark-gray area indicates the instrument compliance limit. The brighter area indicates the temporal region under analysis in this work, for sake of clarity split  in four phases according with the V$_G$ modulation and labeled with the blue circles. (b) A detailed view of the temporal behavior in the region indicated by the red circle in the upper panel  of (a).}
\label{fig6}
\end{figure}\subsubsection{Sample C} 
This is under many aspects the most interesting sample (Fig. \ref{fig6}). Even for this low-carrier density sample, the QW is close to complete filling before the FPP, as shown by the very minor resistance decrease at positive bias. This suggests a general statement, that samples with lower carrier densities are not characterized by an emptier QW with respect to higher carrier density samples, but by a ``smaller" QW, hosting a smaller number of bound states. We observe that in the retention state, the system is insulating both at zero and at negative $V_G$. This means that the FPP persistently changes a highly metallic sample into an insulator, which can only temporarily recover its pristine metallic character during a positive bias. 

As a peculiar feature of sample \textbf{C}, at difference to other samples, at every successive positive pulse of the waveform (``Phase 2"), the same peculiar exponentially damped resistivity increase typical of the FPP takes place. This implies that the QW is filled until the edge, and above, each time a pulse of capacitively induced electrons with surface density $\sigma_{\!_{IND}}$ is provided.
\subsubsection{Erasing procedure}
Let’s now switch to the second part of this paper, by addressing the issue of how can the retention state be erased. This is not only crucial for future hypothetical applications. It is a condition for performing further experimental work, since a phenomenon that can be seen only once in the life of a sample would certainly not allow one any systematic investigation. Biscaras et al.\cite{biscaras_limit_2014} reported a full recovery of the pristine state during a warm-up of the sample towards room temperature. Our experiments lead to slightly different results, suggesting that the behavior reported in \cite{biscaras_limit_2014}, at least, does not represent the general case.
\begin{figure}
\includegraphics[width=0.48\textwidth]{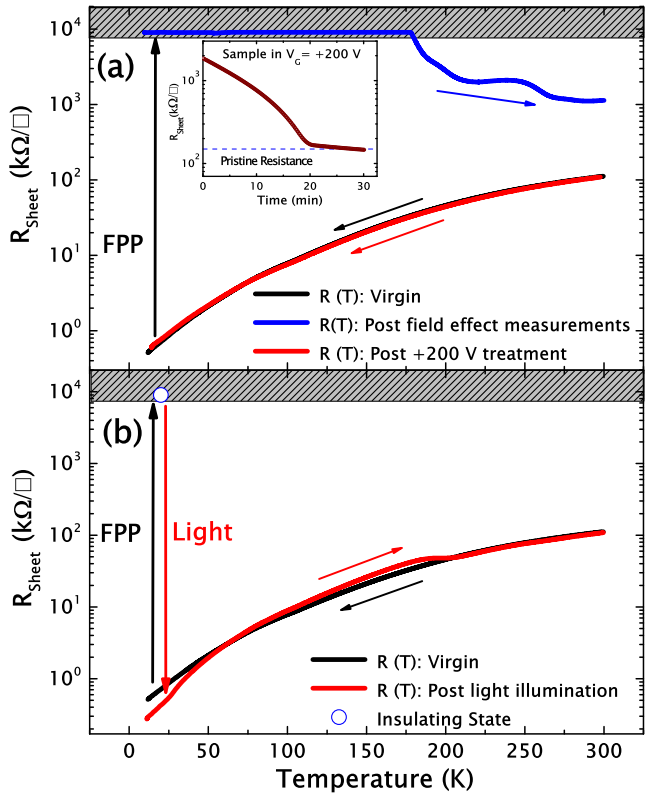} 
\caption{Comparison of subsequent R$_{Sheet}$(T) curves for the sample \textbf{C} in different conditions:
pristine state, retention state and during the erasing procedure. (a) The erasing procedure is performed applying a positive BG voltage of 200V and heating up the system to room temperature. In the inset the temporal evolution of the R$_{Sheet}$ once at 300K. (b) A different erasing procedure is carried out by illuminating   for few seconds the sample with a 625 nm LED at $V_G=0$ . The system is then warmed up to room temperature. }
\label{fig7}
\end{figure}
A number of subsequent experiments are summarized in Fig. \ref{fig7}a for sample C. The sample was first cooled to 20K (black R(T) curve). The FPP performed at 20K brought the sample into its immeasurably resistive retention state, as shown by the blue R(T) curve. During the warm-up phase, the sample  ``attempted" to recover the pristine resistance, through a number of steps taking place at specific activation temperatures. In spite of such partial recovery, it preserved a room temperature resistance an order of magnitude higher than the pristine value. In order to recover the pristine state, we attempted a field-annealing procedure, by keeping the sample under positive voltage for about 30 minutes at room temperature. A slow resistance decrease was obtained, as shown in the inset. At the end of such procedure, the samples indeed recovered the pristine properties, as shown by the red R(T) curve. The same procedure successfully applied to all samples, although a much longer time (above 10 h) was needed to reset sample \textbf{A} and sample \textbf{B}, as shown in \cite{mynote5}. 

On the base of the above-described model of the oxide interface retention state, we assume that the positive voltage, coupled to the room temperature thermal energy, induces a slow creep of trapped charges from the FG states into the control gate, across the whole thickness of the substrate. This procedure is reminiscent of the reset procedure of current electrically erasable programmable read-only memories (E$^2$PROM) devices, also based on FG MOSFET technology. At difference to our samples, characterized by a macroscopically thick control gate barrier, E$^2$PROMs can be reset instantaneously by a control gate polarization cycle, thanks to their ultra-thin oxide gate layer.
\begin{figure}
\includegraphics[width=0.48\textwidth]{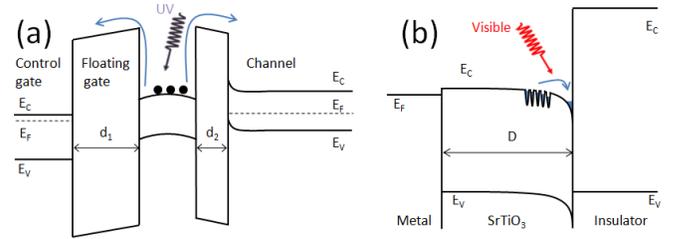} 
\caption{Comparison between the light-induced erasing mechanism for (a) a FG MOSFET device with UV radiation and (b) a BG oxide 2DEG with visible radiation.}
\label{fig8}
\end{figure}

An instantaneous resetting method for our samples was finally mutuated from previous FG EPROM technology(Fig. \ref{fig7}b). For EPROMs, which did not allow for an instantaneous electrical reset, UV light was used to bring the device out of the metastable retention state and back to the pristine state. The high energy of UV photons was necessary for the electrons in the FG to exceed the oxide barrier energy, as sketched in Fig. \ref{fig8}. Due to the small energy barriers confining our trapped electrons in their localized states, as suggested by our sketch reported in Fig. \ref{fig8}b, visible red light (625 nm) was employed in our experiment  at $V_G=0$. The sample was first switched to the highly resistive state by a positive polarization pulse and then switched back by a total photon dose of about $5\times10^{14}$ photons/cm$^2$ delivered by a LED source in about 3-5 seconds (Fig. \ref{fig8}b). This colossal and unprecedented, light-induced insulator-to-metal transition results in a sudden switch with a resistive ratio $R_{OFF}/R_{ON}>10^4$ at 20K. The physics of with phenomenon and the intensity of the effect are totally new with respect to standard photoconductivity effects,  \cite{di_gennaro_photoresponse_2015, di_gennaro_persistent_2013,rastogi_photoconducting_2012} as demonstrated in detail in \cite{mynote5}.

After illumination, a R(T) curve was taken during warm-up. The sample reproduced the curve of the pristine samples, except for two minor features: a photoconductivity effect in the first part of the warm-up curve, and a typical feature at about 200 K, often seen in LAO/STO warm-up plots \cite{siemons_experimental_2007,seri_thermally_2013}.

The reported data and their interpretation in terms of band diagrams pose strong limits to our capability to truly probe the enriched portion of the LAO/STO phase diagrams by a back-gate bias. Data collected on three different samples show that the sheet conductance in the nominal  ``carrier-enriched" state  is equal (sample \textbf{C}) or smaller (samples \textbf{A}, \textbf{B})  than in the pristine state, for all cycles of our BG square waveform, except for the first. Even during the first cycle, i.e. soon after the FPP, the sheet conductivity exceeds the pristine one by no more that 10\% (sample \textbf{A}) or by much less (samples \textbf{B} and \textbf{C}). This is related to the fact that the available, empty, confined 2DEG states, marked in gray in Fig. \ref{fig3}b, are very few in pristine samples with respect to the occupied ones. As shown in \citep{scopigno_phase_2016} and discussed in Appendix A,  such states should not exist at all in an  ``ideal" system where no localized states are present.  The presence of a space charge accumulated in the LAS will allow instead for a non-null second derivative of the potential profile, thus effectively increasing the quantum well edge $W$ shown in Fig. \ref{fig3}b. The space charge in the LAS, induced during the FPP, plays therefore an essential role in making the 2DEG tunable in enrichment mode during the following cycles. These considerations also suggest that a limited occupancy of the LAS might be present also in pristine samples and play a role, by inducing the formation of a weak space charge, in determining the actual LAO/STO band diagram and thus in confining the 2DEG.    

Our data support a scenario in which shallow localized acceptor states are present in proximity of the 2DEG, as theoretically suggested in \cite{scopigno_phase_2016}, and in which a dynamical exchange of carriers between the mobile and trapped states controllably takes place under field effect and light. Any experiment trying to identify the chemical origin of the trap states would go beyond the scope of this work. Our present data only allow to state that the trap states lie in the $STO$ gap and that their depth is in-between several times room temperature kT values (i.e. about 0.1 eV) and our red light photon energy (i.e. 2 eV). Interestingly, the presence of trap states with energy lying about 0.3 eV below the STO conduction band has been demonstrated in STO single crystals by optical spectroscopy\cite{yamakata_behavior_2015}.  The origin of such trap states might be possibly related to the presence of cation vacancies, or of interstitial hydrogen\cite{salman_direct_2014}. 

\section{Conclusions}
We have investigated the  low-temperature transport properties of back-gated 2DEGs at oxide interfaces. Our batch of analyzed samples included metallic amorphous-LGO/STO interfaces, a system for which the formation of an interfacial 2DEG was never proved before.

We show that our samples can be brought by a positive polarization pulse into a persistent insulating state. The magnitude of the effect is analyzed as a function of the initial carrier density.  In low-carrier-density samples, a non-volatile four-order-of-magnitude resistive switching to the insulating state is obtained. 

This work introduces a new element of complexity to our understanding of LAO/STO-like interfaces, addressing them as systems in which two distinct electron reservoirs, mobile carriers and localized charges, interact electrostatically, while dynamically exchanging electrons under external perturbations. We provided evidence that the colossal resistive switching taking place under field effect can be understood in terms of the formation of a retention state that mimics the analogous state of floating-gate MOSFET devices. By suggesting the use of light as a resetting technique, this analogy finally lead us to the fabrication of a non-volatile 3-terminal device prototype, where a  4-order-of-magnitude resistance switch can be induced by a voltage pulse and erased by a pulse of light.

\appendix
\section{\label{sec:Model}}
\subsection{Assumptions}

The behavior of the three different samples in each of the successive voltage steps can be properly described by a simple electrostatic model, which partially neglects the quantum effects of the backgating potential on the confining QW (as shown by Scopigno et al. \cite{scopigno_phase_2016}) but still provides a clear and reliable understanding of the physical system, consistent with the results already found by Biscaras et al. \cite{biscaras_limit_2014}. The model is based on few reasonable and straightforward assumptions:

\begin{enumerate}[I]
\item In agreement with the qualitative analysis reported above, the charge-neutrality conditions for our samples should be written as $\sigma_{\!_{QW}}$= -$\sigma_{\!_{DON}}$=$\sigma_{\!_{2DEG}}$+$\sigma_{\!_{EMPTY}}$=$\sigma_{\!_{2DEG}}$+$\sigma_{\!_{LAS}}$. Such equation implies, in particular,  that in case all localized acceptor states were empty, all the bound states in the quantum well would be occupied,  realizing the condition $\sigma_{\!_{2DEG}}$=$\sigma_{\!_{QW}}$=-$\sigma_{\!_{DON}}$. In case a nonzero  +$\sigma_{\!_{LAS}}$ density due the occupation of the localized states is present, an equivalent amount of empty bound states $\sigma_{\!_{EMPTY}}$=$\sigma_{\!_{LAS}}$, is also present at the interface.

\item No field-effect-induced $p$ doping of the interface is possible. Therefore the SCDs $\sigma_{\!_{2DEG}}$ and $\sigma_{\!_{LAS}}$ can never become positive.
\item As already first observed by \cite{biscaras_limit_2014}, the FPP effect, i.e. the loss of carriers from the QW carriers taking place for a high enough positive $V_G$ is an irreversible process  at the low temperatures (20K) of our experiment. Reformulated in the language of this paper, this means that the system is  unable to transfer charge back from $\sigma_{\!_{LAS}}$ to $\sigma_{\!_{2DEG}}$, even when backgate-induced changes of the potential profile would make this situation energetically more favorable;
An exception to the former assumption takes place, nevertheless, if the system, during a backgate voltage step, has no way to satisfy charge neutrality conditions by keeping $\sigma_{\!_{LAS}}$ constant. In this case only, due to the large electrostatic monopole energy involved, carriers will be detrapped. $\sigma_{\!_{LAS}}$ will be consequently decreased to such a level that allows the sample neutrality condition to be satisfied. This assumption will be crucial in understanding the response of sample C, as shown in the following;
\item The parallel-plate capacitor model, used to describe the variations of SCD, breaks down when the 2DEG is completely depleted.  In such a case, in fact, no potential can be defined on the upper face of the STO substrate. The system  ``misses one plate" and can simply not be modeled anymore as a capacitor. During any attempt to push the sample in further depletion by an extra negative voltage, the sample reacts to the negative pulse as an open circuit,  and remains basically unperturbed, with no further charge transfer taking place. 
\end{enumerate}

 In order to describe the different SCDs of our different samples (\textbf{A}, \textbf{B}, \textbf{C}) in the relevant quarter-cycles  (1, ..., 6) we will adopt a couple of indexes. We start by writing  the charge neutrality conditions of the overall system (including donor states, interface, LAS and back-gate) that, for a generic sample X = A, B or C, in each of the four quarter-cycles of the $V_G(t)$ period, read:
\begin{equation}\label{eq1}
\sigma^{\!_{X,1}}_{\!_{2DEG}}+\sigma^{\!_{X,1}}_{\!_{LAS}}+\sigma^{\!_{X}}_{\!_{DON}}=0
\end{equation}
\begin{equation}
\sigma^{\!_{X,2}}_{\!_{2DEG}}+\sigma^{\!_{X,2}}_{\!_{LAS}}+\sigma^{\!_{X}}_{\!_{DON}}-\sigma^{\!_{X}}\!_{\!_{IND}}=0
\end{equation}
\begin{equation}\label{eq3}
\sigma^{\!_{X,3}}_{\!_{2DEG}}+\sigma^{\!_{X,3}}_{\!_{LAS}}+\sigma^{\!_{X}}_{\!_{DON}}=0
\end{equation}
\begin{equation}\label{eq4}
\sigma^{\!_{X,4}}_{\!_{2DEG}}+\sigma^{\!_{X,4}}_{\!_{LAS}}+\sigma^{\!_{X}}_{\!_{DON}}+\sigma^{\!_{X}}_{\!_{IND}}=0
\end{equation}

\subsection{Some numbers}
The sample is initially in a pristine state, which depends both on the deposition conditions and, possibly, on previous interaction with the environment (e.g., exposure to light). Field-effect measurement estimates of the low-temperature carrier densities of our sample were employed \cite{mynote4}. As shown in \cite{ueno_field_2006} on $La_{0.33}Sr_{0.67}FeO_3$, field-effect carrier density measurements can yield good agreement with Hall measurement also in oxides.  We estimate :
\begin{subequations}
\begin{align}
\sigma^{\!_{A,P}}_{\!_{2DEG}}\simeq 6\cdot10^{13}e~cm^{-2} \\ 
\sigma^{\!_{B,P}}_{\!_{2DEG}}\simeq 4\cdot10^{13}e~cm^{-2} \\ 
\sigma^{\!_{C,P}}_{\!_{2DEG}}\simeq 1\cdot10^{13}e~cm^{-2} 
\end{align}
\end{subequations}
where the P index stay for pristine.
The low-temperature Hall carrier density of sample C, our most important sample, was directly measured as a double-check  \cite{mynote5}, yielding an independently estimation $\sigma^{\!_{C,P}}_{\!_{2DEG}}\simeq 1.1\cdot10^{13}e~cm^{-2}$ in perfect agreement with previous data. The numerical considerations reported below should not be intended as a claim of real quantitive estimation of the carrier densities of the different samples during the cycle. Still, they clearly allow to capture the phenomenological response of the three samples to the successive phases of the voltage cycle.

$\sigma_{\!_{EMPTY}}$ can be estimated by first measuring the maximum positive $V_G$ value leading to a reversible R($V_G$) plot, similarly to the analysis made in \cite{biscaras_limit_2014}, and then estimating the induced SCD according to \cite{hirose_electric_2015}. In the case of sample \textbf{A}, the filling of the QW took place for $V_G$=40V, similarly to the sample analyzed by \cite{biscaras_limit_2014}. In the case of sample \textbf{C}, the filling of the QW took place at about $V_G$=15V. This allows us making the order of magnitude estimates 
\begin{subequations}
\begin{align}
\sigma^{\!_{A,1}}_{\!_{EMPTY}}\simeq 8\cdot10^{12}e~cm^{-2}  \\ 
\sigma^{\!_{C,1}}_{\!_{EMPTY}}\simeq 3\cdot10^{12}e~cm^{-2}
\end{align}
\end{subequations}
Finally, based on an analytic expression for the field and temperature dependence of the $SrTiO_3$ dielectric constant \cite{hirose_electric_2015}, a direct estimate we can estimate of $\sigma_{\!_{IND}}$ can be made:
\begin{subequations}
\begin{align}
\sigma_{\!_{IND}}= 2.5\cdot10^{13}e~cm^{-2} \quad @200V \nonumber
\end{align}
\end{subequations}

\subsection{Analysis of the sample behavior under field effect}
On the base of the former assumptions and definitions, let’s now address the response of all the samples in all the first 6 steps, or quarter-cycles. The response of the system is determined, for any sample $X$ and any step $n$, by the occupation of the QW, $\sigma^{\!_{X,n}}_{\!_{2DEG}}$ and of the LAS, $\sigma^{\!_{X,n}}_{\!_{LAS}}$.
\subsubsection{1$^{st}$ quarter-cycle}
We can generically write that, before any backgate voltage is applied:
\begin{subequations}
\begin{align}
\sigma^{\!_{X,1}}_{\!_{2DEG}}&= \sigma^{\!_{X,P}}_{\!_{2DEG}} \\ 
\sigma^{\!_{X,1}}_{\!_{LAS}}&= \sigma^{\!_{X,P}}_{\!_{LAS}} \\
\sigma^{\!_{X,1}}_{\!_{EMPTY}}&= \sigma^{\!_{X,1}}_{\!_{LAS}}
\end{align}
\end{subequations}
 where the index P indicated that the sample is in the  pristine condition.
\subsubsection{2$^{nd}$ quarter-cycle}
When a positive back-gate voltage is applied, a SCD = $\sigma_{\!_{IND}}$ will be induced at the interface. The electrons will first fill the empty states, adding a SCD = $\sigma_{\!_{EMPTY}}$  into the quantum well and will then populate the LAS. The new SCD will be:
\begin{subequations}
\begin{align}
\sigma^{\!_{X,2}}_{\!_{2DEG}}&=\sigma^{\!_{X,1}}_{\!_{2DEG}} +\sigma^{\!_{X,1}}_{\!_{EMPTY}}=\sigma_{\!_{QW}} =-\sigma_{\!_{DON}}\\ 
\sigma^{\!_{X,2}}_{\!_{LAS}}&= \sigma^{\!_{X,1}}_{\!_{LAS}} + \left(\sigma_{\!_{IND}} - 
\sigma^{\!_{X,1}}_{\!_{EMPTY}} \right) = \sigma_{\!_{IND}}
\end{align} 
\end{subequations}	
The relative amount of resistance decrease at the beginning of phase 2 depends thus, grossly, on the relative amount of empty states above the Fermi level. 
\subsubsection{3$^{rd}$ quarter-cycle}
When $V_G$ is set back to 0V, the electrons trapped in the LAS are not released back to the 2DEG, according to Assumption III, and the population $\sigma_{\!_{LAS}}$ remains unchanged. By combining this assumption with Eq. \ref{eq3}, we obtain the following equations, valid only for sample \textbf{A} and \textbf{B}, for which $|\sigma_{\!_{DON}}|> |\sigma_{\!_{IND}}|$
\begin{subequations}
\begin{align}
\sigma^{\!_{A/B,3}}_{\!_{2DEG}}&=-\sigma^{\!_{A/B}}_{\!_{DON}} -\sigma^{\!_{A/B,2}}_{\!_{LAS}}=
-\sigma^{\!_{A/B}}_{\!_{DON}}-\sigma_{\!_{IND}} \label{eq10a}\\ 
\sigma^{\!_{A/B,3}}_{\!_{LAS}}&= \sigma^{\!_{A/B,2}}_{\!_{LAS}}=\sigma_{\!_{IND}}
\end{align} 
\end{subequations}	
In reasonable agreement with the previous estimates for $\sigma^{\!_{A,1}}_{\!_{2DEG}}$,  $\sigma^{\!_{B,1}}_{\!_{2DEG}}$, $\sigma^{\!_{A,1}}_{\!_{EMPTY}}$,  $\sigma^{\!_{B,1}}_{\!_{EMPTY}}$ and $\sigma_{\!_{IND}}$, the sheet resistance of samples \textbf{A} and \textbf{B} increases by about a factor 2.
For the case of sample \textbf{C}, $|\sigma_{\!_{DON}}|< |\sigma_{\!_{IND}}|$. Equation \ref{eq10a} can not apply, since it would imply a positive charge density at the interface, in disagreement with assumption IV. It is deduced that, during the decrease of $V_G$ from 200V to 0, the 2DEG will be completely depleted at some point ($\sigma^{\!_{C,3}}_{\!_{2DEG}}=0$). Below such point, according to assumption II, the carrier density can not change sign and remains constant, while the LAS will be depleted, according to assumption IV, to preserve charge neutrality (Eq. \ref{eq3}). As a result:
\begin{subequations}
\begin{align}
\sigma^{\!_{C,3}}_{\!_{2DEG}}&=0 \label{eq11a}\\ 
\sigma^{\!_{C,3}}_{\!_{LAS}}&= -\sigma^{\!_{C}}_{\!_{DON}}
\end{align} 
\end{subequations}
\subsubsection{4$^{th}$ quarter-cycle}
When $V_G$ is set back to $ -–200V$, eq. \ref{eq4} applies. Furthermore, Eq. A9b still applies (assumption III). From the combination of the two, we obtain, for sample \textbf{A}:
\begin{subequations}
\begin{align}
\sigma^{\!_{A,4}}_{\!_{2DEG}}&=-\sigma^{\!_{A}}_{\!_{DON}}-2\sigma_{\!_{IND}} \label{eq12a}\\ 
\sigma^{\!_{A,4}}_{\!_{LAS}}&\simeq \sigma_{\!_{IND}}
\end{align} 
\end{subequations}
Sample \textbf{A} has increased its resistivity by a factor 4 with respect to the pristine state, but it is still conducting. This suggests that  $|\sigma^{\!_{A}}_{\!_{DON}}|\simeq$ 2 $|\sigma_{\!_{IND}}|$ , in reasonable agreement with the previous estimates for $\sigma^{\!_{A,1}}_{\!_{2DEG}}$, $\sigma^{\!_{A,1}}_{\!_{EMPTY}}$, and $\sigma_{\!_{IND}}$.

Let’s now analyze first sample \textbf{C}, which is already in deep depletion mode. According to hypothesis IV, nothing happens and the conditions of the previous step still apply
\begin{subequations}
\begin{align}
\sigma^{\!_{C,3}}_{\!_{2DEG}}&=0 \label{eq13a}\\ 
\sigma^{\!_{C,3}}_{\!_{LAS}}&= -\sigma^{\!_{C,3}}_{\!_{DON}}
\end{align} 
\end{subequations}

Sample \textbf{B} is in an intermediate situation. During the beginning of the negative pulse, is it fully conducting. At $-200V$ it is insulating, while  $\sigma^{\!_{B,4}}_{\!_{LAS}}= -\sigma^{\!_{B}}_{\!_{DON}}$ still applies. On the base of the previous estimates for $\sigma^{\!_{B,1}}_{\!_{2DEG}}$,  and $\sigma_{\!_{IND}}$. it can be deduced that the sample becomes insulating at some voltage value $V_G$=$V^*$ pretty close to $-200V$. At such voltage, the condition -$\sigma^{\!_{C,4}}_{\!_{2DEG}}$ = $\sigma^{\!_{B}}_{\!_{DON}}$ - 2$\sigma_{\!_{IND}}(V^*)=0$, applies. For increasing $V_G$, the charges remain frozen, similarly to the previous case of sample \textbf{C}. The conditions are:
\begin{subequations}
\begin{align}
\sigma^{\!_{B,4}}_{\!_{2DEG}}&=0 \label{eq14a}\\ 
\sigma^{\!_{B,4}}_{\!_{LAS}}&= -\sigma^{\!_{B}}_{\!_{DON}}
\end{align} 
\end{subequations}
\subsubsection{1$^{st}$ quarter-cycle of the second cycle (5$^{th}$ step)}
When $V_G$ is set back to 0V, step 5, $\sigma_{\!_{LAS}}$ remains unchanged, maintaining the same value as in step 4. From Eq. \ref{eq1} and from the previous values $\sigma^{\!_{X,4}}_{\!_{LAS}}$, we can deduce the state of the different samples:
\begin{itemize}
\item Sample \textbf{A}
\begin{subequations}
\begin{align}
\sigma^{\!_{A,5}}_{\!_{2DEG}}&=-\sigma^{\!_{A}}_{\!_{DON}}-\sigma_{\!_{IND}} \label{eq15a}\\ 
\sigma^{\!_{A,5}}_{\!_{LAS}}&\simeq \sigma_{\!_{IND}}
\end{align} 
\end{subequations}
similarly to the third quarter-cycle. 
\item Sample \textbf{B}\\ 
The population of the LAS remains the same, and under Eq. \ref{eq1} the same conditions as sample \textbf{A} apply:
\begin{subequations}
\begin{align}
\sigma^{\!_{B,5}}_{\!_{2DEG}}&=-\sigma^{\!_{B}}_{\!_{DON}}-\sigma_{\!_{IND}} \label{eq16a}\\ 
\sigma^{\!_{B,5}}_{\!_{LAS}}&\simeq \sigma_{\!_{IND}}
\end{align} 
\end{subequations}		
\item Sample \textbf{C}\\ 
It remains in depletion mode, in the same conditions as in step 3 and 4.
\begin{subequations}
\begin{align}
\sigma^{\!_{C,5}}_{\!_{2DEG}}&= 0 \label{eq17a}\\ 
\sigma^{\!_{C,5}}_{\!_{LAS}}&= -\sigma^{\!_{C}}_{\!_{DON}}
\end{align} 
\end{subequations}	
\end{itemize}

At the end of this 5$^{th}$ quarter-cycle we see that the resistivity measured in all the three samples does not recover the value attained in the 3$^{rd}$ one, despite our model predicts $\sigma_{\!_{2DEG}}$ and $\sigma_{\!_{LAS}}$ the be the same. This is particularly relevant for samples \textbf{A} and \textbf{B}.
We expect this difference to be due to a residual polarization of STO substrate, in agreement with \cite{bell_dominant_2009}. Such hypothesis that will be explored in a future work and cannot be described by our present simplified model.


\subsubsection{2$^{nd}$ quarter-cycle of the second cycle (6$^{th}$ step)}
During the next positive polarization (step 6 and the following second quarter-cycles: 10, etc.) samples \textbf{A} and \textbf{B} show two unexpected behaviors, not shared by sample \textbf{C}: the R(t) curve has both a different (positive) second derivative and a much higher value than the one in step 2.
The first phenomenon can be explained observing that the charges induced by the positive $V_G$ in samples \textbf{A} and \textbf{B} will grossly fill the available state in the QW, since $\sigma^{\!_{A/B,6}}_{\!_{EMPTY}}$ = $\sigma^{\!_{A/B,6}}_{\!_{LAS}}\simeq$ $ \sigma_{\!_{IND}}$, but are not expected to flow out of the QW.
On the other hand, the latter phenomenon is not easily described within our framework and its origin will not be discussed in this work.

Sample \textbf{C} instead replicates at every successive positive polarization the FPP effect shown in Fig.6b, i.e., an exponentially damped recovery towards higher resistance values.

It is in fact recognized that in the case of sample \textbf{C}, $\sigma^{\!_{C,6}}_{\!_{EMPTY}}$ = $\sigma^{\!_{C,6}}_{\!_{LAS}}\simeq$ $-\sigma^{\!_{C}}_{\!_{DON}}<$ $ \sigma_{\!_{IND}}$, therefore the induced charges will exceed, at every time, the available states in the QW,  thus replicating, at every time,  the  FPP effect.

%

\end{document}